\documentclass{article}
\usepackage{spconf,amsmath,graphicx}
\usepackage{makecell}
\usepackage[table]{xcolor}
\usepackage{color}
\usepackage{multirow}
\graphicspath{{images/}}
\title{Siamese Capsule Network for End-to-End Speaker\\Recognition In The Wild}

\name{Amirhossein Hajavi, Ali Etemad}
\address{Department of ECE \& Ingenuity Labs Research Institute\\Queen's University, Kingston\\ 
\small{\texttt{\{a.hajavi, ali.etemad\}@queensu.ca}}}

\begin{document}

\maketitle

\begin{abstract}
We propose an end-to-end deep model for speaker verification in the wild. Our model uses thin-ResNet for extracting speaker embeddings from utterances and a Siamese capsule network and dynamic routing as the Back-end to calculate a similarity score between the embeddings. We conduct a series of experiments and comparisons on our model to state-of-the-art solutions, showing that our model outperforms all the other models using substantially less amount of training data.  We also perform additional experiments to study the impact of different speaker embeddings on the Siamese capsule network. We show that the best performance is achieved by using embeddings obtained directly from the feature aggregation module of the Front-end and passing them to higher capsules using dynamic routing. 
\end{abstract}

\begin{keywords}
Deep Speaker Recognition, End-to-End Speaker Recognition, Siamese Networks, Capsules
\end{keywords}
\section{Introduction}
Speaker verification models are comprised of two main parts. The \textit{Front-end} component which encodes an utterance into fixed sized embedding vectors \cite{dehak2010front}, and the \textit{Back-end} component which measures the similarity of two given vectors in the form of a similarity score \cite{kenny2010bayesian}. Most commonly in previous studies, the two components of the speaker verification system are two separate processes: the Front-end process of embedding extraction through a Deep Neural Network (DNN) pipeline, and the Back-end conducted via a non-trainable secondary module, typically cosine similarity.

Deep learning has emerged as a successful tool for speaker recognition in the past years. In this context, using DNN as a substitute for different components of conventional techniques such as i-Vector \cite{dehak2010front, kenny2010bayesian} has been subject to many studies \cite{snyder_TDNN_2019,wang_RNN_2019,kim_deep_2019, hajavi2019, hajavi2020}. Replacing the Front-end component of the i-Vector/PLDA with DNN models in order to extract speech representations from utterances is one of the common techniques used in recent literature for both speaker identification and verification \cite{kim_deep_2019, hajavi2019, hajavi2020}. Architectures such as Time-Delay Neural Networks (TDNN) \cite{snyder_TDNN_2019}, Recurrent Neural Networks (RNN) \cite{wang_RNN_2019}, and Convolutional Neural Networks (CNN) \cite{kim_deep_2019, hajavi2019, hajavi2020} are some examples of DNN models used in the recent studies.

The most common technique used for the Back-end component in DNN-based speaker recognition is the cosine distance. Despite the advancement of DNN models in extracting learnt speech representations, a very limited number of studies have used sophisticated DNN models to replace cosine distance in the Back-end component. The neural networks used as the Back-end component have typically consisted of simple Multi-Layer Perceptrons (MLP) \cite{rohdin2018mlp_siamese, zhang2016end_siamese, sriskandaraja2018deep_siamese}, whereas more sophisticated methods, such as Siamese networks, have shown to be useful in other application areas \cite{neill2018siamese_capsule}. Siamese networks have been utilized to measure the similarity level between two feature vectors\cite{siamese_book}. The network takes the two vectors as inputs and produces an output score indicating the degree of similarity between the two vectors. Different forms of Siamese networks have proven to be successful in many speech-related tasks such as language detection \cite{shon_2018_siamese_language}, domain adaptation \cite{rozenberg2020siamese_domain}, and speech representation learning \cite{riad2018sampling_siamese_speech}.  

While Siamese networks that incorporate MLPs often successfully learn simple input-output transformations, they fail to consider the part-whole relations within the feature vectors.
% While Siamese networks are trained to detect the differences between two feature vectors, it should be noted that the information used in such networks 
% During the extraction of the information needed for detecting the differences, some parts of information have a higher decisive impact while accumulated with other as a whole. 
On the other hand, capsule networks along with routing mechanisms, have been designed to detect part-whole relations, and have recently emerged \cite{sabour2017dynamic_capsule} as an upcoming approach in deep learning with impressive results obtained in image recognition \cite{kosiorek2019stacked_capsule}, speech emotion recognition \cite{wu2019emotion_capsule}, keyword detection \cite{xiong2019keyword_capsule}, and brain-computer interfaces \cite{zhang2019capsule}. The integration of pose matrices inside of capsules enables capsule networks to support different instantiating parameters such as deformation and orientation. Capsules also use different routing mechanisms such as dynamic routing \cite{sabour2017dynamic_capsule} or EM routing \cite{hinton2018_capsule_em} to capture the part-whole relationships or whole-part relationships \cite{kosiorek2019stacked_capsule} in the input features. 

In this paper we propose a speaker recognition model with a Back-end Siamese capsule network for text-independent speaker verification. Our model can be integrated with any Front-end speaker representation learning model, resulting in a thoroughly end-to-end pipeline. Our contributions in this paper can be summarized as follows. (\textbf{1}) We propose a novel Siamese network using capsules for speaker recognition. We integrate our proposed model with a state-of-the-art Front-end DNN to perform speaker verification. (\textbf{2}) We train the end-to-end pipeline with Voxceleb1 to perform speaker recognition in the wild. Our results outperform that of other solutions and set a new state-of-the-art. (\textbf{3}) We show that despite the fact that the dataset used for training our model (Voxceleb1) is much smaller than Voxceleb2 which is used by some other studies, our model still achieves superior results.

The rest of the paper is as follows. In Section \ref{sec:related}, some of the related works around Siamese networks, end-to-end speaker recognition, and capsules will be discussed. Section \ref{sec:model} will explain the architecture and the details of our proposed model. In section \ref{sec:experiments} the conducted experiments and results are presented. And finally Section \ref{sec:conclusion} concludes the paper and presents suggestions for future work.
%Half a page

\section{Related Work}
\label{sec:related}
\subsection{Siamese Networks for Speaker Recognition}
Deep learning approaches for speaker recognition have gained a lot of attention given the advancements in computational capacity and availability of large in-the-wild datasets \cite{nagrani2020voxceleb, mclaren2016speakers}. A large number of studies using DNN models for speaker embedding extraction have been performed in the past few years. Most prominent studies have used various CNN architectures such as ResNet \cite{xie_utterance-level_2019, hajavi2019, hajavi2020} to achieve effective speaker embeddings from spectral representations of the utterances. Other successful models such as X-Vectors \cite{snyder_TDNN_2019} have used TDNN to obtain reliable speaker embeddings using MFCC features.

The majority of DNN models used in speaker recognition take a single utterance as input and provide a fixed-size vector as the speaker embedding for the utterance. Another process is then used to calculate the similarity of the two embeddings obtained from an \textit{enrolment utterance} and a \textit{test utterance} for speaker verification. The most commonly used technique among the recent studies for calculating the similarity score is the cosine similarity (as shown in Equation \ref{eq:cosine_distance}). In the few cases where a similarity score is calculated using a DNN model \cite{rohdin2018mlp_siamese, zhang2016end_siamese, sriskandaraja2018deep_siamese, zhang_2019end_siamese}, the performance has not been comparable to the state-of-the-art.
\begin{equation}
    Score(V_1,V_2)= \frac{V_1^T\times V_2^{}}{|V_1||V_2|}
\label{eq:cosine_distance}
\end{equation}

\subsection{Capsule Networks}
With the emergence of capsule networks \cite{sabour2017dynamic_capsule}, there has been considerable advancements in learning deep representations of data \cite{kosiorek2019stacked_capsule,wu2019emotion_capsule,xiong2019keyword_capsule,zhang2019capsule}. Taking advantage of routing mechanisms such as dynamic routing has enabled capsules to capture part-whole relations. However, to the best of our knowledge, capsule networks have not yet been explored for speaker recognition purposes.

On the topic of using capsules within a Siamese network architecture, a number of papers have explored this strategy in areas outside of speaker recognition. For example, a Siamese capsule network was proposed in \cite{neill2018siamese_capsule}, which used primary capsules to capture the facial parts from CNN embeddings of the input pair of face images. A higher level capsule used routed information from the primary capsules via dynamic routing to construct part-whole relationships. The representations acquired from capsules were then transformed to a secondary latent space and the final similarity score was calculated using a non-linear combination of these representations. 

The work done in \cite{wu2020siamese_capsule} utilized Siamese capsule networks as a tool for calculating the similarity of two short sequences of text via embeddings obtained from Bidirectional Gated Recurrent Units (BGRU). Employing a similar approach as the Siamese capsule network proposed for face recognition, the embeddings were first passed through primary capsules to identify parts of the text (a representation of words and phrases). The higher capsules then used the information from the primary capsules via a routing mechanism to formulate a representation for the whole text.

As the mentioned studies suggest, the use of capsule networks (either as a stand-alone model or as part of a Siamese network) has shown promising results. However there has been no studies on the use of Siamese capsule networks for speaker recognition. In this work we aim to introduce a novel network based on this type of architecture to perform speaker verification using audio signals. 
%Half a page

\section{Proposed Network}
\label{sec:model}

We propose an architecture based on a Siamese network with a Back-end that utilizes capsules for directly measuring a similarity score for speaker verification. Through the following sub-sections, we describe the different components of our model. An overview of the model is presented in Figure \ref{fig:model}.

% \noindent \textbf{Front-end:} 
\subsection{Front-end} 
\begin{figure}[t]
    \begin{center}
    \includegraphics[width=0.9\columnwidth]{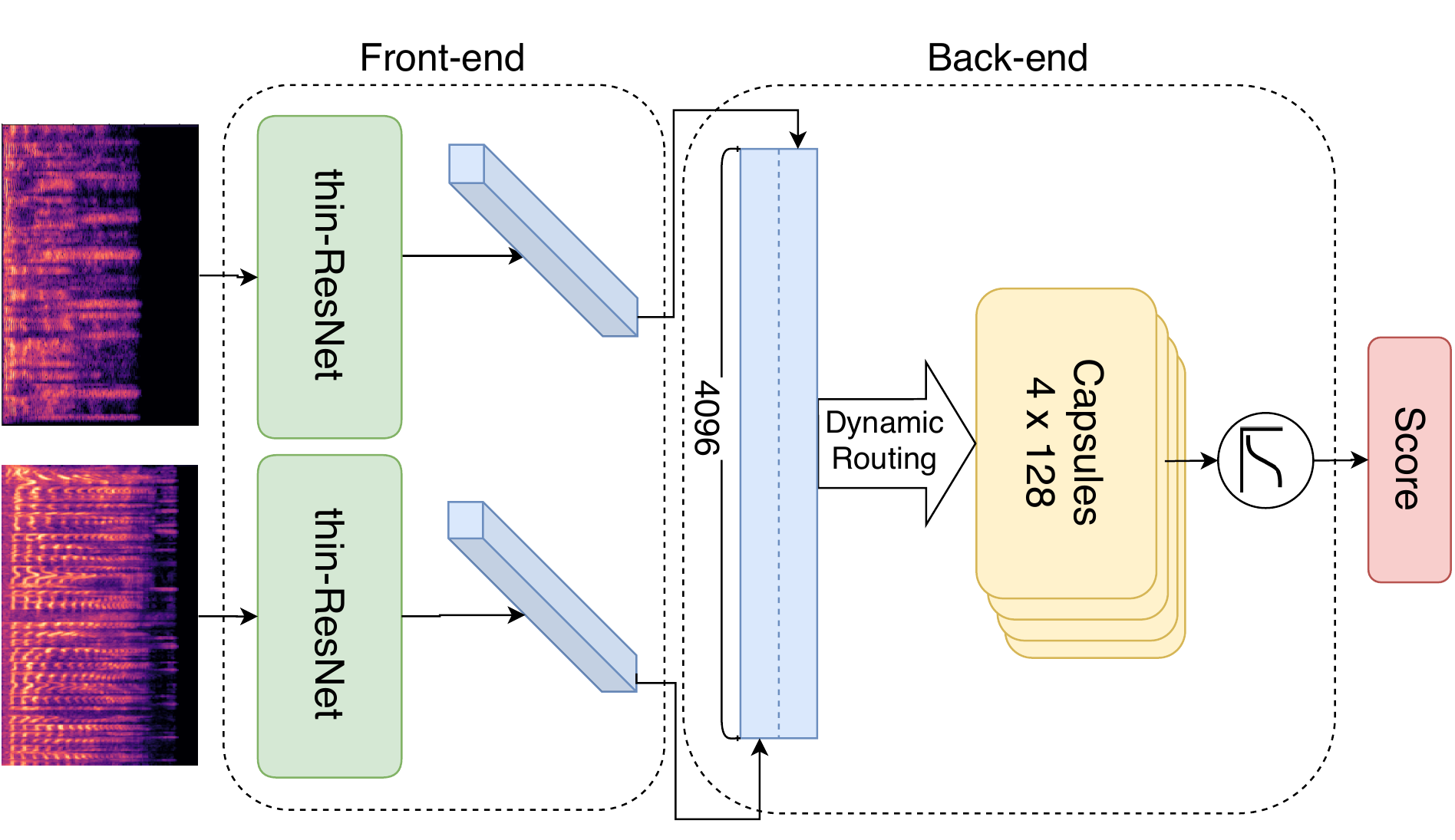}
    \end{center}
    \caption{The architecture of our proposed end-to-end speaker recognition model with Siamese capsule networks.}
\label{fig:model}
\end{figure}

We utilize a model with Resnet-based architecture as the Front-end component of our end-to-end DNN model. In order to better be able to show the impact of the Siamese capsule network, the front-end component as depicted in Figure \ref{fig:model}, shares the majority of its architectural details with the model proposed in \cite{xie_utterance-level_2019}. The model is a modified version of ResNet34, namely thin-ResNet34, which utilizes 34 convolution layers in the main body of the model. It also uses an effective feature aggregation method, namely GhostVlad \cite{xie_utterance-level_2019}. The original model, uses a final fully connected (FC) layer to transform the features into a latent space and the final embedding is scaled down to 512 elemental vectors which are later used for verification via cosine distance. In our paper, we opt to remove the FC layer in order to preserve the maximum information possible from the embedding vectors provided by the GhostVlad pooling mechanism.

The thin-ResNet model operates on the enrollment utterance and the test utterance using a identical set of weights. This results in two vectors of size 4096 for each of the utterances. These vectors are then paired together in matrices with a dimension of $(4096\times 2)$ in the way that each index of the embedding vector of enrollment utterance is paired with the same index in the embedding vector of the test utterance. This helps to better compare the embeddings later on in the Back-end component of our network.

% \noindent \textbf{Back-end:} 
\subsection{Back-end}
\label{sec: model sub: Back-end}
The Back-end component of the DNN model (see Figure \ref{fig:model}) used in this paper consists of only higher capsules as we opt not to use the primary capsules. The operations inside primary capsule layers include a non-linear convolutions and the squashing operation. We aim to compare each index of the enrollment utterance embedding with the same index of the test utterance embedding and the non-linear effect of the convolution layer and the squashing operation inside the primary capsule prevents this correlation. Therefore by skipping the operations of the primary capsule, the embeddings are directly passed to the higher capsules. Each tuple $(v_{1_i},v_{2_i})$, where $v_{1_i}$ is the $i^{th}$ index in the enrollment embedding, and $v_{2_i}$ is the same index in the test embedding, is considered a single \textit{part} while extracting the \textit{part-whole} relations using dynamic routing. 

We utilize 4 higher capsules in the Back-end model. The number of capsules is found empirically and may differ with respect to the diversity and complexity of the set of speakers. We also use dynamic routing \cite{sabour2017dynamic_capsule} with 3 iterations as the routing mechanism between the embedding tuples and the capsules. In this mechanism, embedding tuples are first normalised using the L2 normalization (see Equation \ref{eq:l2}). The contribution of each tuple $V_i$ in the higher capsule $C_j$ is then determined by a secondary coefficient $c_{ij}$ through performing routing softmax (see Equation \ref{eq:dynamic_softmax}). The value $p_{ij}$ is calculated through multiple iterations using gradients obtained from the final loss function. Lastly, the representations from the higher capsules are then aggregated and passed to a Sigmoid function. The model is trained through binary classification and the final score calculated by the Sigmoid function is presented as the similarity score between the enrollment utterance and the test utterance.
\begin{equation}
    V_i = \frac{V_i}{||V_i||+\epsilon}
    \label{eq:l2}
\end{equation}
\begin{equation}
     c_{ij}= \frac{exp(p_{ij})}{\sum\limits_{k} exp(p_{ik})}
    \label{eq:dynamic_softmax}
\end{equation}

\begin{table*}[!t]
  \caption{The results for evaluation of our Siamese capsule network in comparison to several benchmark models.} %proposed in \cite{xie_utterance-level_2019, chung2018_triplet_loss, cai2018analysis}.}
  \label{tab:results_full}
\footnotesize
  \centering
  \begin{tabular}{l l l l l}
    \hline
      & \textbf{Model} & \textbf{Loss} & \textbf{Train set } & \textbf{EER\% } \\

    \hline\hline
    & & & & \\[-.3cm] 
    
    Nagrani et al. \cite{nagrani2020voxceleb} & I-Vector + PLDA & -- & VoxCeleb1 &  8.80 \\%[.1cm]
    
    Cai et al. \cite{cai2018analysis} & ResNet34 + SAP & A-softmax + PLDA  & VoxCeleb1 & 4.40 \\%[.1cm]
    
    Cai et al. \cite{cai2018analysis} & ResNet34 + LDE & A-softmax + PLDA   & VoxCeleb1 & 4.48 \\%[.1cm]
    
    Chung et al. \cite{chung2018_triplet_loss} & ResNet50 + TAP & Triplet Loss & VoxCeleb2 & 4.19 \\%[.1cm]
    
    Hajavi et al. \cite{hajavi2019} & UtterIdNet + TDV & Softmax  & VoxCeleb2 & 4.26\\%[.1cm]
    
    Okabe et al. \cite{okabe2018attentive} & TDNN (X-Vector) + TAP & Softmax  & VoxCeleb1 & 3.85 \\%[.1cm]
    
    Xie et al. \cite{xie_utterance-level_2019} & Thin-ResNet34 + GhostVlad & Softmax & VoxCeleb2 & 3.22 \\%[.1cm]  
    % \hline 
    \textbf{Ours} & Thin-ResNet34 + Siamese Capsule & Binary Cross-entropy & VoxCeleb1 & \textbf{3.14}\\%[.1cm]
    \hline
  \end{tabular}
\end{table*}

\begin{table}[t]
\begin{center}
\caption{The results for evaluation of Siamese capsule networks with different architectures, using embeddings from different layers of the Front-end model. FC: the embedding from the last FC layer of the model; Aggreg.: the embedding from the output of GhostVlad aggregation module.}
\label{tab:ablation}
\footnotesize
\begin{tabular}{c  c  c  c  c}
\hline
Layer & Dimension & No. Caps. & Primary Caps. & EER \\ 
\hline \hline
FC & 512 & 2 & No & 3.86 \\ 
FC & 512 & 4 & No & 3.65 \\ 
FC & 512 & 6 & No & 3.63 \\ \hline
Aggreg. & 4096 & 2 & No & 3.18 \\ 
Aggreg. & 4096 & 4 & No & \textbf{3.14} \\ 
Aggreg. & 4096 & 6 & No & 3.16 \\ \hline
Aggreg. & 4096 & 2 & Yes & 4.06 \\ 
Aggreg. & 4096 & 4 & Yes & 3.83 \\ 
Aggreg. & 4096 & 6 & Yes & 3.90 \\ 
\hline
\end{tabular}%
\end{center}
\end{table}

% \noindent \textbf{Implementation:}
\subsection{Implementation}
Our Siamese capsule network utilizes four higher capsules with a capsule dimension of 128. Each capsule receives information from 4096 tuples through dynamic routing. The number of trainable parameters are approximately $8.3$M parameters. The high number of trainable parameters increases the probability of over-fitting. To address this issue, we opt to train the model using random selection of utterances from various speakers. 

Training of the model is done using a similar approach to that of the triplet loss. For each step of the training, three utterances are selected, two utterances form the same speaker and the other spoken by a different speaker. The same speaker utterances are selected form all the utterances of a speaker using a uniform random distribution without replacement. Then a third utterance is selected from the utterances of a random speaker, reducing the chance of the same triplets being selected again further in training. 

We use Adam optimizer for training our Siamese capsule network. The Front-end model remains frozen during training in order to isolate the source of any changes in performance to the impact of the capsule network only. The learning rate is initially set to 0.01, but is changed with  cyclical learning rate pattern \cite{smith2017cyclical} to ensure optimal convergence. For the hardware, we use a single Titan RTX GPU for training. For batch size, 64 utterance triplets are selected at each step of the training. 
% A page and Half - TWO pages

\section{Experiment Setup and Results}
\label{sec:experiments}

\subsection{Dataset} 
% \textbf{Dataset:} 
In this paper, the VoxCeleb dataset is used for both training and evaluation. The training set consists of approximately 148k utterances spoken by 1,211 speakers. Using the random selection of utterances for triplets, the scale of possible triplet combinations for training adds up to $148,000\times122^2$, where the number 122 is the average utterance count for a speaker in VoxCeleb1 dataset. Hence, selecting VoxCeleb1 dataset with less number of speakers and utterances compared to VoxCeleb2, helps with managing the number of combination.

% \noindent \textbf{Performance:} 
\subsection{Performance}
Table \ref{tab:results_full} presents the results of our experiments. In this table the performance of the Siamese capsule network with respect to the benchmark models of thin-ResNet+GhostVlad, ResNet34+SAP, and ResNet34+LDE is presented. The comparison illustrates that our model achieves an EER of 3.14\%, outperforming all the benchmark models. We also compare the performance of our model with respect to the amount of data needed for training. 
While some models \cite{hajavi2019, xie_utterance-level_2019, chung2018_triplet_loss} use the VoxCeleb2 dataset which contains more than 1M utterances for training, our model utilizes substantially less amount of data, and yet outperforms these models. This may be due to the random selection of utterance triplets which increases the number of training samples provided by the VoxCeleb1 dataset.

We also evaluate the performance of our model using different number of capsules in the architecture. We also test the embeddings from two different layers of the Front-end model and the effect of using primary capsules immediately on the embeddings. Table \ref{tab:ablation} includes the results of our experiments. As shown in the results, the best performance is achieved using four capsules in the architecture. Also the embeddings obtained directly from the GhostVlad aggregation module lead to a considerably better performance compared to the bottleneck features collected from the fully connected layer. The effect of using primary capsules on the embeddings leads to a lower performance. This is in compliance with the argument made earlier (see Section \ref{sec: model sub: Back-end}) that the non-linear operations performed in primary capsules prevent the model from collecting more decisive information from embeddings.

In the end, we should point out that while a number of other works have attempted to use CNN-based Siamese architectures for speaker recognition in the past \cite{zhang2016end_siamese, zhang_2019end_siamese}, they have generally achieved less competitive results compared to the approach of using cosine distance on the obtained embeddings. However, the use of a Siamese capsule network in our model shows improvement over the conventional approach, which can indicate the viability of such architectures for speaker verification in the wild.

\section{Conclusion and Future Work}
\label{sec:conclusion}
In this paper a novel Siamese network using capsules and dynamic routing was proposed for speaker verification in the wild. Our end-to-end pipeline used thin-ResNet as its Front-end component for speech representation learning, while capsules were used in its Back-end to extract part-whole relations of the embeddings later used to calculate the similarity score between two representations. Experiments on the Voxceleb test set illustrated that our model outperforms the other benchmarks by obtaining an EER of 3.14\% and sets a new state-of-the-art. Our model can be trained using substantially less amount of training data to reach the desired performance by random selection of utterance triplets. As a possible future route we intend to extend our experiments of the Siamese capsule network on utterances from various domains such as environments and noise levels.

\section{Acknowledgements}  
The authors would like to thank IMRSV Data Labs for their support of this work and also acknowledge the Natural Sciences and Engineering Research Council of Canada (NSERC) for supporting this research (grant no.: CRDPJ 533919-18).

\small
\bibliographystyle{IEEEtran}
\bibliography{refs}

\end{document}